\documentclass[epj,nopacs]{svjour}
\usepackage{graphicx}
\usepackage{multirow}
\begin{document}
\title{Measurement of light mesons at RHIC by the PHENIX experiment}
\author{M. Naglis (for the PHENIX Collaboration)\inst{}
\thanks{\email{maxim.naglis@weizmann.ac.il}}%
}                     
%
%
\institute{Weizmann Institute of Science, Rehovot 76100, Israel}
\date{Received: date / Revised version: date}
%
\abstract{
The PHENIX experiment at RHIC has measured a variety of light neutral mesons ($\pi^{0}$,
K$_{S}^{0}$, $\eta$, $\omega$, $\eta^{\prime}$, $\phi$) via multi-particle decay channels over a wide
range of transverse momentum. A review of the recent results on the production rates of light
mesons in p+p and their nuclear modification factors in d+Au, Cu+Cu and Au+Au collisions at different
energies is presented.
\PACS{
      {PACS-key}{discribing text of that key}   \and
      {PACS-key}{discribing text of that key}
     } 
} 
\maketitle
\section{Introduction}
\label{intro}
The suppression of high-$p_T$ light mesons in ultra-relativistic heavy ion collisions, compared to
expectations from scaled p+p results, is one of the most interesting findings 
at RHIC. Particles with high transverse momentum are believed to be produced by fragmentation of
hard-scattered partons in the earliest stage of the collision. In nucleus-nucleus collisions,
the scattered hard partons lose their energy while traversing the subsequently produced hot and
dense matter leading to the phenomenon of jet-quenching. Medium-induced effects on
high-$p_T$ particle production can be quantified with the nuclear modification factor:

\begin{equation}
\label{formula1}
  R_{AA}(p_T)=\frac{d^2N_{AA}/dydp_T}{(\langle N_{coll}\rangle /\sigma_{pp}^{inel}) \times
  d^2\sigma_{pp}/dydp_T},  
\end{equation}
where $d^2N_{AA}/dydp_T$ is the differential yield per event in nucleus-nucleus collisions,
$\langle N_{coll}\rangle$ is the number of binary nucleon-nucleon collisions averaged over the
impact parameter range of the corresponding centrality, and $\sigma_{pp}^{inel}$ and
$d^2\sigma_{pp}/dydp_T$ are the total and differential cross sections for inelastic p+p collisions,
respectively. 
In the absence of medium-induced effects, the yield of high-$p_T$ particles is expected to scale
with $\langle N_{coll}\rangle$, resulting in  R$_{AA}=1$ at high-$p_T$.
Contrary to the suppression of mesons, baryons are found to be enhanced in central Au+Au
collisions at RHIC at intermediate transverse momenta ($1.5<p_T<4.5$~GeV/$c$) \cite{Adler:2003kg}.
The difference in the behavior of mesons and baryons is still not well understood. 

Precise knowledge of meson and baryon yields and their ratios as a function of transverse momentum and
in different collision systems, provides a mean to investigate the dependence of hadron
formation on particle mass and quark-flavor composition, and the mechanisms of jet production and
quenching. The PHENIX experiment excels in measuring light neutral mesons via multi-particle decay
channels. This paper presents a review of the light meson data obtained in p+p and heavy-ion
collisions at different energies. 

\section{Experimental setup and data analysis}
\label{detector}

The PHENIX detector \cite{Adcox:2003zm} at RHIC (Relativistic Heavy Ion Collider) has been
designed to measure a broad variety of signals from heavy ion collisions. 
PHENIX is capable of measuring electrons, photons and  hadrons in a large dynamic range.  
The tracking of charged particles and measurements of their momentum are accomplished by
the high-resolution multi-wire proportional Drift Chambers (DC) and the first layer of Pad
Chambers (PC1). The typical value of the momentum resolution is $\sigma(p_T)/p_T \approx
1.0\%p_T~\oplus~1.1\%$. The energy of photons and electrons is measured by highly segmented
lead-glass (PbGl) and lead-scintillator (PbSc) Electro Magnetic Calorimeters (EMCal) which have
an energy resolution of $\sigma_{PbGl}(E)/E \approx 5.9\%/\sqrt{E}~\oplus~0.8\%$ and
$\sigma_{PbSc}(E)/E \approx 8.1\%/\sqrt{E}~\oplus~2.1\%$, respectively. Charged hadron
identification is based on the particle mass calculated from the momentum and the time-of-flight
information derived from the TOF or EMCal subsystems and the Beam Beam Counters (BBC's).
Reliable pion-kaon separation can be achieved in the $p_T$ range $0.3-2.5$~GeV/$c$ and
$0.3-1.0$~GeV/$c$ using the TOF and the PbSc part of the EMCal, respectively. The Zero Degree
Calorimeters (ZDC's) and BBC's are dedicated subsystems that determine the collision vertex and event
centrality and also provide the minimum bias (MB) interaction trigger. 
The results of data presented in this paper are based on data samples  accumulated during p+p, d+Au,
Cu+Cu and Au+Au collisions at $\sqrt{s_{NN}}=200$~GeV, p+p and Cu+Cu collisions at
$\sqrt{s_{NN}}=62.4$~GeV, and Cu+Cu collisions at $\sqrt{s_{NN}}=22.4$~GeV.
The MB trigger used for p+p, d+Au and Cu+Cu collisions requires a coincidence between the BBC's with at
least one hit in each BBC arm. 
For Au+Au collisions the MB trigger requires a coincidence between the BBC's and ZDC's  with
at least two hits in each BBC arm, at least one neutron detected in each ZDC arm.
Besides, PHENIX utilizes a hardware high-$p_T$ photon/electron trigger (ERT) based on the analog 
sum of the signals of adjacent 4$\times$4 EMCal towers in coincidence with the MB trigger condition.
All events used in the analysis are required to have the collision vertex position along the beam axis within
30~cm of the geometrical center of PHENIX.

\begin{table*}[ht!]
\centering
\caption{Summary of analyzed decay channels and data samples.
In Au+Au (p+p) collisions, the measurements of the $\phi$ meson were performed using two different analysis
techniques referred to in text as ''two kaons PID'' and ''no PID''(''one kaon PID'' and ''no PID''), respectively. 
}
\label{tab:1}       
\begin{tabular}{lcccccc}
\hline\noalign{\smallskip}
Data sample &$\sqrt{s_{NN}}$~(GeV)           &$\int Ldt$& Trigger & Decay mode & $p_T$~(GeV/$c$) & Reference\\
\noalign{\smallskip}\hline\noalign{\smallskip}
Run-2 $Au+Au$ &200& 24~$\mu$b$^{-1}$ & MB & $\eta \rightarrow \gamma\gamma$&2.25$-$9.5&\cite{Adler:2006bv}\\
\noalign{\smallskip}\hline\noalign{\smallskip}
\multirow{5}*{Run-3 $p+p$}
&\multirow{5}*{200}&\multirow{5}*{0.35~pb$^{-1}$}& ERT & $K_S^0 \rightarrow \pi^0\pi^0$ &2.5$-$6.5&\cite{Ryabov:2007zz}\\
&&& MB & $\eta \rightarrow \gamma\gamma$&2.75$-$11.0&\cite{Adler:2006bv}\\
&&&ERT & $\eta \rightarrow \pi^0\pi^+\pi^-$&3.0$-$8.0&\cite{Adler:2006bv}\\
&&& ERT & $\omega \rightarrow \pi^0\pi^+\pi^-$ &2.75$-$9.25&\cite{Adler:2006hy}\\
&&& ERT & $\omega \rightarrow \pi^0\gamma$&2.5$-$6.5&\cite{Adler:2006hy}\\
\noalign{\smallskip}\hline\noalign{\smallskip}
\multirow{6}*{Run-3 $d+Au$}
&\multirow{6}*{200}&\multirow{6}*{2.74~nb$^{-1}$}& ERT & $K_S^0 \rightarrow \pi^0\pi^0$ &3.5$-$8.5&\cite{Ryabov:2007zz}\\
&&& MB & $\eta \rightarrow \gamma\gamma$&2.25$-$11.0&\cite{Adler:2006bv}\\
&&& ERT & $\eta \rightarrow \pi^0\pi^+\pi^-$&5.0$-$8.0&\cite{Adler:2006bv}\\
&&& ERT & $\omega \rightarrow \pi^0\pi^+\pi^-$ &3.5$-$9.0&\cite{Adler:2006hy}\\
&&& ERT & $\omega \rightarrow \pi^0\gamma$&3.0$-$7.0&\cite{Adler:2006hy}\\
&&& MB & $\phi \rightarrow K^+K^-$&1.45-5.1&\cite{Ryabov:2008zz}\\
\noalign{\smallskip}\hline\noalign{\smallskip}
\multirow{4}*{Run-4 $Au+Au$}
&\multirow{4}*{200}&\multirow{3}*{241~$\mu$b$^{-1}$}& MB & $\pi^0 \rightarrow \gamma\gamma$ &1.25$-$19.0&\cite{Adare:2008qa}\\
&&& MB & $\omega \rightarrow \pi^0\gamma$ &4.0$-$9.0&\cite{Ryabov:2007zz}\\
&&& MB & $\phi \rightarrow K^+K^-$&0.8$-$4.0&\cite{Pal:2005xy}\\
&&& MB & $\phi \rightarrow K^+K^-$&2.45$-$7.0&-\\
\noalign{\smallskip}\hline\noalign{\smallskip}
\multirow{7}*{Run-5 $p+p$}
&\multirow{7}*{200}&\multirow{7}*{3.78~pb$^{-1}$}& MB/ERT& $\pi^0 \rightarrow \gamma\gamma$& 0.625$-$19.0&\cite{Adare:2007dg}\\
&&& ERT & $K_S^0 \rightarrow \pi^0\pi^0$ &2.25$-$10.0&\cite{Ryabov:2007zz}\\
&&& ERT & $\omega \rightarrow \pi^0\pi^+\pi^-$&2.25$-$13.0&\cite{Ryabov:2007zz}\\
&&& ERT & $\omega \rightarrow \pi^0\gamma$ &2.5$-$11.0&\cite{Ryabov:2007zz}\\
&&& ERT & $\eta^{\prime} \rightarrow \eta\pi^+\pi^-$&3.25$-$10.75&-\\
&&& MB & $\phi \rightarrow K^+K^-$&0.9$-$4.5&-\\
&&& MB & $\phi \rightarrow K^+K^-$&1.3$-$7.0&-\\
\noalign{\smallskip}\hline\noalign{\smallskip}
\multirow{3}*{Run-5 $Cu+Cu$}
&200&3.06~nb$^{-1}$& MB/ERT & $\pi^0 \rightarrow \gamma\gamma$&1.25.0$-$11.0&\cite{Adare:2008cx}\\
&62.4&0.19~nb$^{-1}$& MB & $\pi^0 \rightarrow \gamma\gamma$&1.25$-$7.5&\cite{Adare:2008cx}\\
&22.4&2.7~$\mu$b$^{-1}$& MB & $\pi^0 \rightarrow \gamma\gamma$&1.25$-$4.75&\cite{Adare:2008cx}\\
\noalign{\smallskip}\hline\noalign{\smallskip}
Run-6 $p+p$ &62.4&0.1~pb$^{-1}$& MB/ERT & $\pi^0 \rightarrow \gamma\gamma$&0.625$-$6.75&\cite{Adare:2008cx}\\
\noalign{\smallskip}\hline
\end{tabular}
\end{table*}
A summary of the  decay channels studied and data samples analyzed is given in Table~\ref{tab:1}.
The procedure used to reconstruct the invariant mass spectra of
$\pi^0$, K$_{S}^{0}$, $\eta$, $\eta^{\prime}$, $\omega$ mesons is as follows. 
All photon clusters measured in the EMCal in a given event are paired to form the invariant mass
distribution. Pairs with invariant mass within two standard deviations from the $\pi^0$ ($\eta$)
mass are considered as $\pi^0$ ($\eta$) candidates. Raw $\pi^0$ ($\eta$) yields are determined by
subtracting the yield of combinatorial pairs estimated by pairing photon clusters from different events.
Then, the selected $\pi^0$ ($\eta$) candidates are combined between themselves, or with other photon clusters, or
with charged tracks, according to the decay modes listed in Table~\ref{tab:1}.
The invariant mass spectra of the $\phi$ meson are constructed using three different techniques.
The first does not require track identification in the final state and assumes
that all tracks are kaons (''no PID''). The second requires identification of only one kaon in the TOF
subsystem (''one kaon PID''). In the third technique both kaons are identified in the TOF or EMCal
subsystems (''two kaons PID''). Kaon candidates with opposite charge are combined to form the invariant
mass distribution. Mass distributions obtained by the procedures described above
contain both the signal and an inherent combinatorial background. 
Meson raw yields were extracted by simultaneously fitting the signal and background, or
by integrating the spectra in the vicinity of the meson masses after subtraction of the combinatorial
background, estimated using an event-mixing technique. Corrections to the raw yields for the limited
detector acceptance and resolution, reconstruction and trigger efficiency, multiplicity effects and
various analysis cuts are determined from the full single-particle Monte Carlo simulation and
analysis of the data. 
\section{Results}
\label{results}
\subsection{Particle spectra}
\label{spectra}
\begin{figure}
\centering
\includegraphics[width=0.98\columnwidth,height=0.44\textheight]{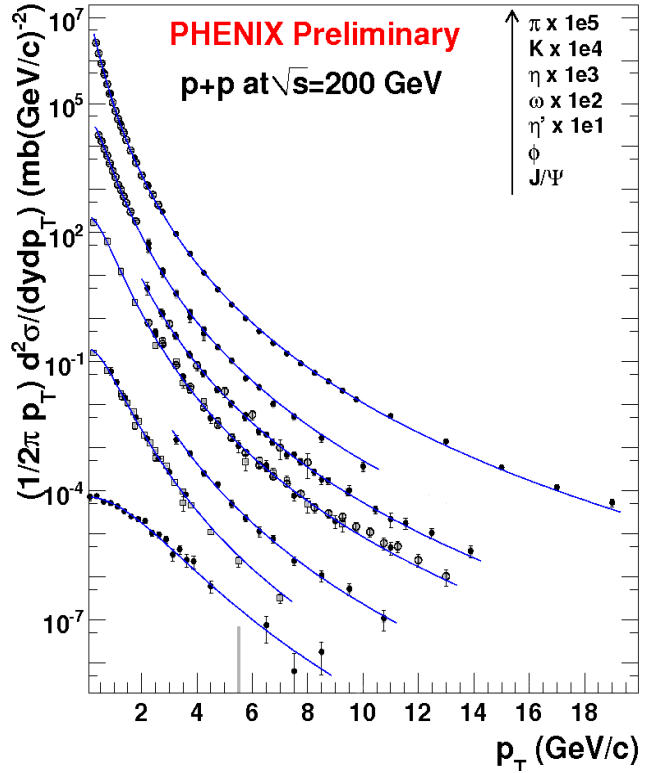} 
\caption{Compilation of PHENIX results on meson production in p+p collisions at
$\sqrt{s}=200$~GeV. The solid lines are just to guide the eye. Charged kaon and pion data are
taken from \cite{Adler:2006xd}, J/$\Psi$ data are taken from \cite{Adare:2006ns}.} 
\label{fig:omega_k_eta}
\end{figure}

The study of inclusive particle production in nucleon - nucleon collisions is a rich source of
information on the fragmentation properties of partons in perturbative QCD and hadronization
mechanisms. The PHENIX experiment has measured a variety of mesons in p+p collisions at
$\sqrt{s}=200$~GeV. These data in comparison to existing leading- and next-to-leading order
perturbative QCD calculation can be used to constrain the model parameters. A compilation of published
and preliminary results on meson production in p+p collisions is presented in Fig.~\ref{fig:omega_k_eta}. 
The different symbols for a given meson correspond to different decay channels
or different analysis techniques. The third set of points from the bottom represents the first
measurement of $\eta^{\prime} \rightarrow \eta\pi^+\pi^-$ at RHIC at high-$p_T$.
The solid lines shown in the figure are just to guide the eye.

The details of the K$_S^0$, $\eta$ and $\omega$ measurements in p+p, d+Au and Au+Au collisions
at $\sqrt{s_{NN}}=200$~GeV can be found elsewhere \cite{Adler:2006hy,Adler:2006bv,Ryabov:2007zz}. 
Good agreement was obtained in the measurements of the same meson via two different decay
channels, $\omega \rightarrow \pi^0\pi^+\pi^-$ or  $\omega \rightarrow \pi^0\gamma$ and $\eta
\rightarrow \pi^0\pi^+\pi^-$ or $\eta \rightarrow \gamma\gamma$.
Analyses of MB and ERT data samples collected in Run-3
and Run-5 gave consistent results. 

\begin{figure}
\centering
  \includegraphics[width=0.98\columnwidth,height=0.27\textheight]{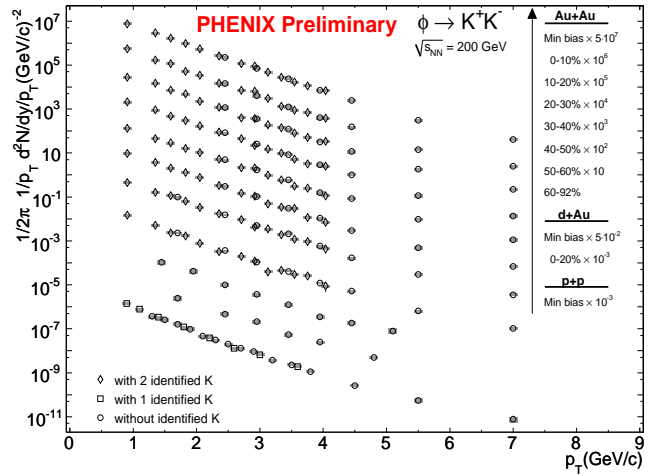}
\caption{Invariant $p_T$ spectra of $\phi$ meson measured in p+p, d+Au and Au+Au collisions at
$\sqrt{s_{NN}}=200$~GeV.}
\label{fig:phi_spectra}       
\end{figure}

The $\phi$ meson invariant $p_T$ spectra measured in p+p, d+Au and Au+Au collisions at
$\sqrt{s_{NN}}=200$~GeV are shown in Fig.~\ref{fig:phi_spectra}. 
As compared to the previously reported Au+Au results \cite{Pal:2005xy} obtained with the ''two
kaons PID'' approach, the new ''no PID'' analysis extends the range of the measurements  to higher
$p_T$ up to 7~GeV/$c$. In turn, the new ''one kaon PID'' analysis of p+p data extends the range of the ''no
PID'' measurements reported in \cite{Ryabov:2008zz} to lower $p_T$ down to 0.9~GeV/$c$. In spite
of the very different sources of systematic uncertainties, the results of the ''no PID'' analysis
are in good agreement with  the results of ''one kaon PID'' and ''two kaons PID'' in
p+p and Au+Au collisions, respectively. The combined p+p results of the ''one kaon PID'' and ''no
PID'' analyses constitute a new p+p reference for $\phi$ meson, which supersedes the one
used previously \cite{Pal:2005xy} having both a wider $p_T$ range and smaller error bars.

\subsection{Particle ratios}
\label{ratios}
Non-identical particle ratios measured at high-$p_T$ are useful to assess the particle
composition of jet fragmentation. In p+p collisions, meson-to-meson ratios ($\eta/\pi^0$,
K/$\pi^0$, $\eta^{\prime}/\pi^0$, $\omega/\pi^0$, $\phi/\pi^0$, $\phi$/K) derived using a
parameterization of the $\pi^0$ invariant $p_T$ spectra \cite{Adler:2003pb} exhibit a universal
flat behavior above $p_T=2.5$~GeV/$c$ \cite{Ryabov:2008zz} indicating that the measured transverse
momentum spectra for all light mesons have the same slope at high-$p_T$. Similarly to p+p results,
$\eta/\pi^0$, K$_S^0/\pi^0$ and $\omega/\pi^0$ ratios in d+Au collisions are also found to be flat
in the same $p_T$ range. Results of the fits to a constant performed in the $p_T$ region above
2.5~GeV/$c$ for $\eta/\pi^0$, K$_S^0/\pi^0$ and $\omega/\pi^0$ ratios are presented in Table~\ref{tab:2}.
As one can see, the results obtained in p+p and d+Au collisions are consistent within errors for the same
ratio, indicating that the production of light mesons at $p_T>2.5$~GeV/$c$ is not affected by cold nuclear
matter. 

\begin{table*}
\centering
\caption{Results of the fits to a constant for $\eta/\pi^0$, K$_S^0/\pi^0$ and $\omega/\pi^0$ ratios in
p+p and d+Au collisions at $\sqrt{s_{NN}}=200$~GeV.}
\label{tab:2}       
\begin{tabular}{cccc}
\hline\noalign{\smallskip}
 system & $\eta/\pi^0$ & K$_S^0/\pi^0$ & $\omega/\pi^0$\\
\noalign{\smallskip}\hline\noalign{\smallskip}
$p+p$&0.48 $\pm$ 0.02(stat)$\pm$0.02(syst)&0.45 $\pm$ 0.01(stat)$\pm$0.05(syst)&0.81 $\pm$ 0.03(stat)$\pm$0.07(syst)\\
$d+Au$&0.47 $\pm$ 0.02(stat)$\pm$0.02(syst)&0.58 $\pm$ 0.06(stat)$\pm$0.05(syst)&0.94 $\pm$ 0.08(stat)$\pm$0.12(syst)\\
\noalign{\smallskip}\hline
\end{tabular}
\end{table*}

\subsection{Nuclear modification factors}
\label{factor}
\begin{figure}
\centering
\resizebox{0.98\columnwidth}{!}{%
  \includegraphics{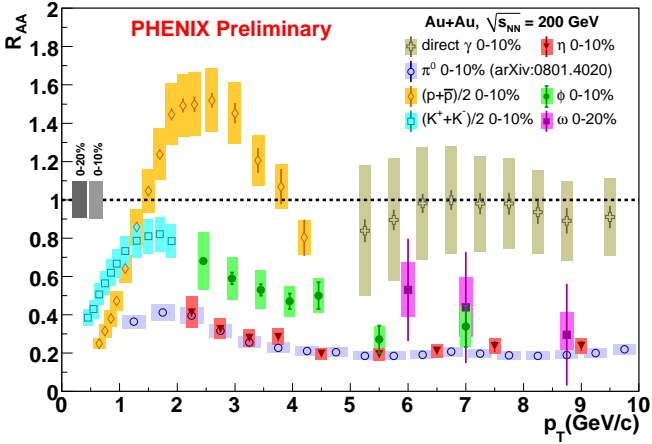}}
\caption{Nuclear modification factor R$_{AA}$ in central Au+Au collisions as a function of $p_T$
for $\pi^0$, ($K^++K^-$)/2, $\eta$, $\omega$, ($p+\bar{p}$)/2, $\phi$  and direct
$\gamma$. Direct $\gamma$ data are taken from \cite{Reygers:2008pq}.}
\label{fig:raa} 
\end{figure}

Using the new p+p reference, $\phi$ mesons's R$_{AA}$  has been derived in central Au+Au
collisions in the $p_T$ range of $2.45-7$~GeV/$c$. There is an ongoing work to extend this $p_T$ range
towards low-$p_T$. 
Fig.~\ref{fig:raa} shows a comparison of the R$_{AA}$ for $\phi$ mesons, other mesons ($\pi^0$,
$\eta$, K, $\omega$), protons and direct photons. Neutral pions and $\eta$ mesons follow the
same suppression pattern over the entire $p_T$ range. The R$_{AA}$ of $\phi$ mesons exhibits less
suppression than $\eta$ and $\pi^0$ in the $p_T$ range of $2.45<p_T<4.5$~GeV/$c$. 
At higher $p_T$ ($>5$~GeV/$c$) the amount of suppression of $\phi$ and $\omega$ could be
similar to that of $\pi^0$ and $\eta$ mesons. A more conclusive statement is not possible due to
the large errors of the measurements. 
The similarity between the suppression patterns of different mesons at high-$p_T$
favors the production of these mesons via jet fragmentation outside the hot and dense medium
created in the collision. It is not clear whether the R$_{AA}$ of charged kaons follow or not the
trend of $\phi$ mesons, since the present measurements have no overlap in $p_T$. Interestingly,
that R$_{AA}$ of protons which are known to be enhanced in the intermediate $p_T$ range
\cite{Adler:2003kg}, reach a value below unity at $p_T \approx 4.2$~GeV/$c$. Further investigations
are required for understanding the influence of quark flavor composition on the suppression pattern.

\subsection{Onset of jet quenching}
\label{onset}
\begin{figure}
\centering
  \includegraphics[width=0.98\columnwidth,height=0.26\textheight]{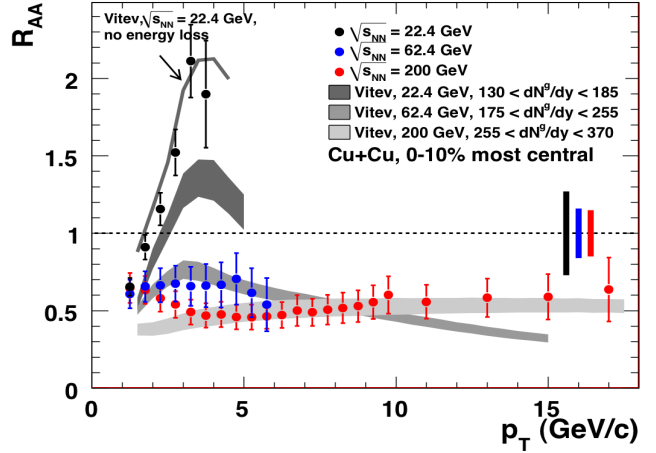}
\caption{R$_{AA}$ of $\pi^0$ as a function of transfer momentum in central Cu+Cu collisions at $\sqrt{s_{NN}}=22.4$~GeV, 62.4~GeV and
200~GeV shown in black, blue and red, respectively. The shaded bands represent parton energy
loss calculations \cite{Adare:2008cx}.}
\label{fig:onset}       
\end{figure}

In contrast to RHIC results at $\sqrt{s_{NN}}=130$~GeV and 200~GeV, no suppression of high-$p_T$
hadrons has been observed at SPS energies. This implies that there 
must be an intermediate value of collision energy at which the onset of high-$p_T$ suppression
happens. The results of a beam energy scan performed at RHIC with Cu+Cu collisions at
$\sqrt{s_{NN}}=22.4$~GeV, 62.4~GeV and 200~GeV, shown in Fig.~\ref{fig:onset}, indicate that
parton energy loss starts to dominate over Cronin enhancement at a center of mass energy somewhere
between 22.4 and 62.4 GeV per nucleon pair \cite{Adare:2008cx}. Numerical calculations based on a
parton energy loss model \cite{Vitev:2005he} describe well the suppression of $\pi^0$ mesons above
$p_T=3$~GeV/$c$ at 62.4~GeV and 200~GeV. Measurements at 22.4~GeV favor the calculations carried out
without taking into account parton energy loss.

\section{Summary and outlook}
The PHENIX experiment at RHIC has performed a set of measurements of light mesons ($\pi^{0}$,
K$_{S}^{0}$, $\eta$, $\omega$, $\eta^{\prime}$, $\phi$) via multi-particle decay channels at
intermediate and  high $p_T$ in different collision systems.
A first measurement of $\eta^{\prime} \rightarrow \eta\pi^+\pi^-$ at RHIC performed by PHENIX
at high-$p_T$ is reported here. It would be very interesting to measure the production rate
of $\eta^{\prime}$ in Au+Au collisions in view of predictions of a possible enhancement as its
mass decreases in the hot and dense matter \cite{Kapusta:1995ww}. 
In p+p and d+Au $p_T$-dependent non-identical meson-to-meson ratios exhibit a universal flat
behavior above $p_T=2.5$~GeV/$c$ demonstrating that $p_T$ spectra of all measured light mesons have
the same slope in this $p_T$ range. Cold nuclear matter does not affect the production 
rates of light mesons at $p_T>2.5$~GeV/$c$ as seen from the matching of meson-to-meson ratios
measured in p+p and d+Au. 
Using the new p+p reference, $\phi$ meson's R$_{AA}$ has been derived in central Au+Au
collisions. 
At intermediate-$p_T$ $\phi$ meson is less suppressed than $\pi^0$ and $\eta$ and 
it exhibits similar suppression level to those of $\pi^0$ and $\eta$ at high-$p_T$.
Parton energy loss calculations reproduce well the $\pi^0$ meson suppression for $p_T>3$~GeV/$c$ in
Cu+Cu collisions at $\sqrt{s_{NN}}=62.4$~GeV and 200~GeV. The onset of jet quenching is between
22.4~GeV and 62.4~GeV.
\begin{acknowledgement}
The author acknowledges support by the Israel Science Foundation, the MINERVA Foundation and the
Nella and Leon Benoziyo Center of High Energy Physics Research.
\end{acknowledgement}


\begin{thebibliography}{}
%
%


\bibitem{Adler:2003kg}
  S.~S.~Adler {\it et al.},
	    Phys.\ Rev.\ Lett.\  {\bf 91}, 172301 (2003).

\bibitem{Adcox:2003zm}
  K.~Adcox {\it et al.},
  Nucl.\ Instrum.\ Meth.\  A {\bf 499} (2003) 469.

\bibitem{Adler:2006xd}
  S.~S.~Adler {\it et al.},
	    Phys.\ Rev.\  C {\bf 74}, 024904 (2006).

\bibitem{Adare:2006ns}
  A.~Adare {\it et al.},
	    Phys.\ Rev.\ Lett.\  {\bf 98}, 232301 (2007).

\bibitem{Adler:2006hy}
  S.~S.~Adler {\it et al.},
   Phys.\ Rev.\  C {\bf 75}, 051902 (2007).

\bibitem{Adler:2006bv}
  S.~S.~Adler {\it et al.},
	Phys.\ Rev.\  C {\bf 75}, 024909 (2007).

\bibitem{Ryabov:2007zz}
  V.~Ryabov (for the PHENIX Collaboration),
	  Int.\ J.\ Mod.\ Phys.\  E {\bf 16}, 1864 (2007).

\bibitem{Pal:2005xy}
  D.~Pal (for the PHENIX Collaboration),
  Nucl.\ Phys.\  A {\bf 774}, 489 (2006).

\bibitem{Ryabov:2008zz}
  V.~Ryabov (for the PHENIX Collaboration),
	  J.\ Phys.\ G {\bf 35}, 044030 (2008).
			
\bibitem{Adler:2003pb}
  S.~S.~Adler {\it et al.},
  Phys.\ Rev.\ Lett.\  {\bf 91}, 241803 (2003).

\bibitem{Adler:2006wg}
  S.~S.~Adler {\it et al.}
	    Phys.\ Rev.\ Lett.\  {\bf 98}, 172302 (2007).

\bibitem{Reygers:2008pq}
  K.~Reygers (for the PHENIX Collaboration),
	    arXiv:0804.4562 [nucl-ex].


\bibitem{Adare:2008cx}
  A.~Adare {\it et al.},
  arXiv:0801.4555 [nucl-ex].

\bibitem{Vitev:2005he}
  I.~Vitev,
	  Phys.\ Lett.\  B {\bf 639}, 38 (2006).

\bibitem{Adare:2007dg}
  A.~Adare {\it et al.},
		  Phys.\ Rev.\  D {\bf 76}, 051106 (2007).

\bibitem{Adare:2008qa}
  A.~Adare {\it et al.},
		  arXiv:0801.4020 [nucl-ex].

\bibitem{Kapusta:1995ww}
  J.~I.~Kapusta {\it et al.},
	  Phys.\ Rev.\  D {\bf 53}, 5028 (1996).

\end{thebibliography}
\end{document}